# Magnetic Cluster Expansion model for random and ordered magnetic face-centered cubic Fe-Ni-Cr alloys

M.Y. Lavrentiev[(1)], J.S. Wróbel[(1)], D. Nguyen-Manh[(1)], S.L. Dudarev[(1)], and M.G. Ganchenkova[(2)]

*(1) CCFE, Culham Science Centre, Abingdon, Oxon, OX14 3DB, United Kingdom*
*(2) Materials Science Department, National Research Nuclear University MEPhI,*
*31 Kashirskoe sh., 115409, Moscow, Russia*

**Abstract**

A Magnetic Cluster Expansion (MCE) model for ternary face-centered cubic Fe-Ni-Cr alloys has been developed using DFT data spanning binary and ternary alloy configurations. Using this MCE model Hamiltonian, we perform Monte Carlo simulations and explore magnetic structures of alloys over the entire range of alloy compositions, considering both random and ordered alloy structures. In random alloys, the removal of magnetic collinearity constraint reduces the total magnetic moment but does not affect the predicted range of compositions where the alloys adopt low temperature ferromagnetic configurations. During alloying of ordered fcc Fe-Ni compounds with Cr, chromium atoms tend to replace nickel rather than iron atoms. Replacement of Ni by Cr in alloys with high iron content increases the Curie temperature of the alloys. This can be explained by strong antiferromagnetic Fe-Cr coupling, similar to that found in bcc Fe-Cr solutions, where the Curie temperature increase, predicted by simulations as a function of Cr concentration, is confirmed by experimental observations.

## I. Introduction

Fe-Cr-Ni based austenitic stainless steels retain high mechanical strength at elevated temperatures, making them attractive structural materials for light water and fast breeder fission reactors [1]. Because of its robustness, austenitic stainless steel 316L(N) was selected as a structural material for ITER [2]. Until now, very few comprehensive theoretical investigations of Fe-Cr-Ni ternary alloy system were performed, owing to the difficulty of treating the interplay between structural order and magnetism in these alloys. Recently, we have developed an *ab initio* parameterized Heisenberg–Landau lattice Hamiltonian-based Magnetic Cluster Expansion (MCE) model for binary fcc Fe–Ni [3]. To describe the high- *and* low-spin magnetic configurations of fcc Fe, terms up to the $8^{th}$ order in atomic magnetic moment were included in the Landau expansion for the on-site terms in the Hamiltonian. Thermodynamic and magnetic properties of the alloys were explored, using configurational and magnetic Monte Carlo simulations, over a broad temperature range extending well over 1000 K. The predicted fcc-bcc coexistence curve, the phase stability of ordered $Fe_3Ni$, FeNi, and $FeNi_3$ intermetallic compounds, and the predicted temperatures of magnetic transitions simulated as functions of alloy compositions were found to agree well with experimental observations. In particular, simulations show that magnetic interactions stabilize fcc phases of binary Fe–Ni alloys. Parameters of the MCE model for Fe-Ni alloys were derived from DFT calculations performed for a large number of representative atomic configurations, as well as from DFT data on pure fcc Ni and Fe. The success of that model, together with the availability of DFT data accumulated in the context of a recent comprehensive *ab initio* investigation of Fe-Ni-Cr alloys [4], makes it possible to extend MCE to ternary fcc Fe-Ni-Cr alloys.

The MCE model for ternary Fe-Ni-Cr alloys is the first example of application of Magnetic Cluster Expansion to a magnetic alloy containing more than two components. The initial parameterization of the Fe-Ni-Cr MCE Hamiltonian, and initial simulations performed using this Hamiltonian, are described in Ref. [4]. Here we describe an improved more accurate MCE model based on a larger DFT database of structures and magnetic configurations. Monte Carlo simulations using the MCE Hamiltonian span both random and ordered alloy structures. The advantages of MCE include the possibility of simulating a broad range of alloy compositions and a large variety of chemical and magnetic configurations. Also, MCE makes it possible to study magnetic properties of both ferro- and antiferromagnetic alloys. This aspect of the model is particularly significant in relation to *fcc* Fe-Ni-Cr alloys, since Ni at low temperature is ferromagnetic, whereas pure fcc Fe and Cr, according to *ab initio* calculations, have vanishingly small magnetic moments.

## II. Magnetic Cluster Expansion model for a Ternary Alloy

MCE has been applied to a variety of binary magnetic alloys, including bcc and fcc Fe-Cr [5] and fcc Fe-Ni [3]. Combining a lattice MCE Hamiltonian model with experimental data on vibrational spectra, we explained the origin of bcc-fcc structural phase transitions in pure Fe, and reproduced the occurrence of fcc γ-loop in the Fe-Cr phase diagram. In the case of Fe-Ni alloys, a phase diagram including both the disordered alloy configurations and ordered FeNi and FeNi$_3$ compounds was derived [3]. A large DFT dataset of atomic structures and magnetic configurations accumulated as a part of a recent investigation of Fe-Ni-Cr alloys [4] now makes it possible extend Magnetic Cluster Expansion treatment of fcc Fe-Ni system to the ternary alloy case. Within MCE formalism [6,7], an alloy configuration is defined by its discrete chemical ($\sigma_i$) *and* continuous magnetic ($\mathbf{M}_i$) atomic degrees of freedom. To simplify applications of MCE to a ternary alloy and reduce the number of model parameters, the ternary alloy MCE Hamiltonian includes only pairwise interatomic interactions. The energy of an arbitrary structural and magnetic alloy configuration $(\{\sigma_i\}, \{\mathbf{M}_i\})$ in an MCE model has the form:

$$
\begin{aligned}
E(\{\sigma_i\}, \{\mathbf{M}_i\}) = & \sum_{ij \in 1NN} I^{(1NN)}_{\sigma_i \sigma_j} + \sum_{ij \in 2NN} I^{(2NN)}_{\sigma_i \sigma_j} + \ldots \\
& + \sum_i A_{\sigma_i} \mathbf{M}_i^2 + \sum_i B_{\sigma_i} \mathbf{M}_i^4 + \sum_i C_{\sigma_i} \mathbf{M}_i^6 + \sum_i D_{\sigma_i} \mathbf{M}_i^8 + \\
& \sum_{ij \in 1NN} Y^{(1NN)}_{\sigma_i \sigma_j} \mathbf{M}_i \cdot \mathbf{M}_j + \sum_{ij \in 2NN} Y^{(2NN)}_{\sigma_i \sigma_j} \mathbf{M}_i \cdot \mathbf{M}_j + \ldots ,
\end{aligned}
\qquad (1)
$$

where $\sigma_i$, $\sigma_j$ = Fe, Cr, or Ni, $\mathbf{M}_i$ is the magnetic moment of atom *i*, and the non-magnetic and magnetic interaction parameters ($I_{ij}$ and $Y_{ij}$, respectively) for each set of neighbours in the lattice are 3×3 matrices defined in the discrete space of atomic species. Parameters *A*, *B*, *C* and *D* in (1) are the Landau coefficients for the quadratic, quartic, 6[th]- and 8[th]-order magnetic self-energy terms, respectively. To make the model consistent with the MCE Hamiltonian for Fe-Ni alloys, the 29 binary fcc Fe-Ni configurations used in fitting the MCE model for fcc Fe-Ni alloys to DFT data [3] were also used in this study. The magnetic Fe-Fe and Ni-Ni interaction parameters are retained from the earlier Fe-Ni MCE parameterization [3], whereas the possibility of varying the Fe-Ni interaction parameters is included in the new fit. In addition to binary Fe-Ni configurations, the new parameterization involves DFT data for the 31 ordered ternary Fe-Cr-Ni structures spanning the entire alloy composition triangle, together with the DFT data on pure elements. List of ternary structures used in the fit is presented in the Supplementary

Materials of Ref. [4]. No Ni-Cr binary alloy configurations were used as input for the fitting procedure, and hence MCE predictions for alloys with low iron content are expected to be less accurate than those for iron-rich alloys. *Ab initio* calculations were performed using the projector augmented wave method implemented in VASP package. Similarly to the binary Fe-Ni case, MCE model Hamiltonian interaction parameters were assumed to extend up to the fourth nearest neighbor in fcc lattice. In total the model involves 24 non-magnetic ($I_{ij}$) and 24 magnetic ($Y_{ij}$) interaction parameters. At the initial stage of fitting, the on-site terms *A*, *B* etc. were fitted using the energy versus magnetic moment curves computed for ferromagnetic pure Fe, Ni, and Cr. For chromium, only the quadratic and quartic terms in the Landau expansion for the energy versus magnetic moment were used, whereas for iron and nickel the on-site Landau expansion was extended to the $8^{th}$ order in magnetic moment. The dependence of the on-site energy terms on local atomic environment was neglected to reduce the number of model parameters. Subsequently, using the procedure described in [3], fitting of interaction terms *I* and *Y* was performed for both energies and magnetic moments on each atom in the simulation cell. DFT and MCE energies of mixing for the structures included in the fit are shown in Figure 1. The average error of the fit for energies is 18 meV/atom. A complete list of interatomic MCE interaction parameters for Fe-Cr-Ni alloys is given in the Table 1. The on-site terms are given in Table 2.

To verify the accuracy of MCE fit for magnetic moments, we selected several special quasi-random structures (SQS) with Fe content close to 70 at. %, Cr content close to 18 at. %, and Ni content close to 12 at. %. The structures comprised 108 atoms each, corresponding to 27 (3×3×3) fcc unit cells. Magnetic moments of atoms in these structures were calculated in the collinear approximation using DFT. Next, MCE simulations were performed in two different ways, with and without imposing the collinearity constraint on the directions of magnetic moments. Table 3 compares results obtained using the two approaches. For MCE simulations performed in the collinear approximation, the two approaches agree well for the four out of five structures investigated here. Once the collinearity requirement was removed, atomic magnetic moments rotated away from their magnetization axis, and the total magnetic moment of the alloy decreased. Also, the moments of individual atomic species decreased. The non-collinear magnetic configurations were found to be more stable than collinear configurations, although the energy gain associated with the relaxation of collinear magnetic states into non-collinear states was relatively small, varying from 3 to 11 meV/atom, which is within the accuracy of the fit. It was therefore not possible to conclude without ambiguity whether the true ground state was collinear or non-collinear.

Most of the structures used in the parameterization of MCE Hamiltonian (1) belonged to the Fe-rich area of the ternary phase diagram, and to Fe-Ni solid solutions. Hence we expect that predictions derived from

MCE simulations should be more accurate for alloys where Fe content exceeds 50 at. %, as well as for alloys where Fe and Ni are the dominant components. Almost all the Monte Carlo simulations were performed using 16384 atom simulation cells (containing 16×16×16 fcc unit cells). Each Monte Carlo run included 80000 attempts to change magnetic moment per atom at the equilibration stage, and the same number of Monte Carlo attempts at the subsequent accumulation stage. As an example of application of MCE to low-temperature magnetic properties of a ternary alloy, as well as another test of accuracy of the MCE fit for magnetic moments, we investigated the dependence of the total magnetic moment of $(Fe_{0.5}Ni_{0.5})_{1-x}Cr_x$ alloys on Cr content. An ordered Fe-Ni alloy with $L1_0$ structure was used as an initial configuration, and Cr content was then increased by replacing equal numbers of Fe and Ni atoms with Cr atoms in two ways: (i) by keeping the structures ordered and supercell small and (ii) by randomly choosing the atoms to be replaced in a large supercell. Figure 2 shows the total magnetic moments predicted by MCE for the alloys formed in this way. Simulations were performed with and without the collinearity constraint. As Cr content increases, magnetization rapidly decreases, resulting in an almost completely nonmagnetic alloy at $x_{Cr} = 0.5$, in agreement with *ab initio* DFT calculations.

## III. Random Fe-Ni-Cr Mixtures

Technologically important Fe-Ni-Cr austenitic steels [1,2,8] are usually produced at high temperatures, 1000° C or higher [9]. At reactor-relevant operating temperatures of over 0.3 $T_m$, where $T_m$ is the melting temperature, and at a high irradiation dose [10], the structure of alloys is close to a completely random solid mixture. For example, in almost all the experimentally investigated binary Fe-Cr, Fe-Ni, and ternary Fe-Ni-Cr alloys [11-18], the absolute magnitude of Warren-Cowley short-range order parameters does not exceed 0.1 for any of the three pairs of elements. This shows that the completely random ternary solid solution approximation provides a good representation of a real alloy.

The search for magnetic ground states spanned the entire range of alloy compositions. The concentration step for each element was 6.25 at. %. Three-stage magnetic quenching was performed, in the temperature interval from T=1000 K to T=1 K (first stage), then down to $10^{-3}$ K (second stage), and finally to $10^{-6}$ K (third stage). Pure fcc Fe and Cr were found to have vanishing total magnetic moments, in agreement with DFT calculations. Experimental studies of coherent Fe precipitates in fcc Cu matrix show that the magnetic ground state of fcc Fe is non-collinear [19,20], whereas for fcc Cr only a non-magnetic ground state was found in DFT calculations [4]. Our MCE-based simulations predict a non-collinear magnetic ground state for Fe, while for fcc Cr collinear antiferromagnetic ground state was found, which is only 6 meV/atom more favourable energetically than a non-magnetic ground state. Pure fcc Ni is predicted to be collinear ferromagnetic, also in agreement with experiment. As a result, random alloy structures with non-vanishing total magnetic moment are predominantly found in the Ni-rich part of the alloy composition triangle. Figure 3 shows the total magnetic moment at T 0 K (ground state) as a function of alloy composition found in simulations performed with and without the collinearity constraint. In both cases the addition of up to 50 at. % of Fe or Cr to pure nickel increases the overall magnetic moment per atom, and alloy remains ferromagnetic. While for Fe-Ni alloys this agrees well with *ab initio* data [4], in the Ni-Cr alloy system a rapid *decrease* of the total magnetic moment was found both in DFT [4] and experimental studies [21], with the total moment vanishing above 20 at. % Cr concentration. This disagreement of MCE predictions with *ab initio* and experimental data likely results from the fact that no Ni-Cr binary structures were used in fitting the MCE model Hamiltonian, and explains why MCE predictions for alloys with low iron content are less accurate than those for iron-rich compounds. At higher concentration of Fe or Cr, the total moment found in MCE simulations decreases rapidly, and alloys become antiferromagnetic once the concentration of Ni drops below 25 at. %. Application of a collinearity constraint leads to the overall increase of the average magnetic moment. The predicted areas in the ternary concentration triangle where the total moment is non-zero, largely coincide irrespectively whether the simulations are performed in the collinear or non-collinear approximation. This includes the Fe-Cr composition line. The occurrence of an interval of concentration where alloys have non-zero total magnetic moment stems from strong antiferromagnetic coupling between Fe and Cr. This gives rise to the non-compensation of the total moment once the iron content exceeds that of chromium. It is instructive to compare Figure 3 with Figure 9 of Ref. [4], which shows magnetic moments of various ordered Fe-Ni-Cr compounds. The pattern of variation of magnetic moment over the composition triangle is similar, but the magnitude of magnetic moment is higher for the ordered stable structures compared to random structures, reaching almost 2 $\mu_B$ at the Fe-Ni composition line.

When analysing the energies of Fe-Ni-Cr alloys, it is important to distinguish between the enthalpy of mixing and the enthalpy of formation. The difference between the two entities stems from the fact that in pure Ni fcc structure has the lowest energy, while in Fe and Cr the bcc phases are energetically more stable. The enthalpy of mixing of *fcc* Fe-Ni-Cr is calculated with respect to the enthalpies of constituting elements, assuming that they *all* have fcc crystal structure. The enthalpy of formation, on the other hand, is calculated with respect to the *lowest energy* crystal structures of the constituting pure elements, which in the case of Fe and Cr are bcc. A comprehensive *ab initio* study of various structures was performed in Ref. [4], and in the current work we use the fcc-bcc energy differences derived there, namely, $E_{fcc}$(Fe)-$E_{bcc}$(Fe) = 82 meV/atom; $E_{fcc}$(Cr)-$E_{bcc}$(Cr) = 405 meV/atom; $E_{fcc}$(Ni)-$E_{bcc}$(Ni) = 96 meV/atom.

Enthalpies of mixing and formation computed for fcc Fe-Ni-Cr alloys at T ≈ 0 K are shown in Figure 4. The mixing enthalpy is negative over the entire range of alloy compositions, with the lowest absolute values corresponding Ni-Cr binary mixtures (note that these values characterise random mixtures only). The enthalpy of formation is minimum near the pure Ni corner of the composition triangle.

At high temperatures, magnetic order vanishes for almost all the alloy compositions already at temperatures close to T=500 K (see Figure 5). Ferromagnetism is retained only in the Ni-rich corner of the composition triangle. This agrees with our previous simulations [3] showing that magnetic order in pure Ni predicted by MCE Hamiltonian based simulations vanishes at 550-600 K (the experimental Curie temperature of nickel is 631 K [22]). It is interesting to note that there is also another region where high-temperature magnetic order persists, namely in random Fe-Cr mixtures with alloy compositions in the range from $Fe_2Cr$ to $Fe_3Cr$ (Figure 5). The reason for the occurrence of high-temperature magnetic order here (as well as large magnetic moment at low temperatures, see Figure 3) is related to the strong first nearest neighbour antiferromagnetic interaction between Fe and Cr (Table 1). This produces an effect similar to the one responsible for the Curie temperature of bcc Fe-Cr alloys being maximum at 6 at. % Cr [23], with an important difference since in the case of fcc alloys, ferromagnetism emerges in a mixture of two *antiferromagnetic* metals. The occurrence of ferromagnetic order was also noted in DFT studies [4].

### IV. Ordered Fe-Ni-Cr Structures

Magnetic properties of several ordered Fe-Ni and Fe-Ni-Cr compounds were investigated using MCE-based Monte Carlo simulations. The phase diagram of binary Fe-Ni alloys involves two, or possibly three, ordered stoichiometric compounds, namely FeNi with $L1_0$ structure, $FeNi_3$ and $Fe_3Ni$ with $L1_2$ structure. Whereas $FeNi_3$ is a well-known compound and FeNi (tetrataenite) is found in meteorites [24-26], $Fe_3Ni$ is an assumed compound since it is less stable, compared to random Fe-Ni alloys, than the two other compounds.

Having completed the investigation of binary Fe-Ni alloys [3], we now pose a question about how the addition of chromium influences their energy and magnetic properties. For example, it is desirable to clarify which of the two elements, Fe or Ni, is more readily replaced by chromium. To answer this question, we performed Monte Carlo simulations of all the three above stoichiometric compounds with chromium atoms replacing either Fe or Ni, or both Fe and Ni in equal proportion. The fcc lattice sites where Cr atoms replaced Fe or Ni atoms were chosen at random. Figures 6 (a-c) show the low-temperature enthalpy of the three compounds plotted as a function of Cr content. Cr concentration varied from 0 to 25 at. % for $FeNi_3$ and $Fe_3Ni$, and from 0 to 50 at. % for FeNi. In all the three cases, chromium atoms clearly prefer Ni sites for replacement, with the enthalpy difference being as high as 50 meV/atom. The bias associated with the preferential replacement of Ni by Cr, rather than Fe by Cr, can be explained by the strong Fe-Cr antiferromagnetic interaction in the first nearest neighbor configuration and by the larger magnetic moment of chromium compared to nickel (note that in Hamiltonian (1) the energy of magnetic interaction is a sum of products of interaction parameters and scalar products of magnetic moments themselves, and not the moment unit vectors).

It is reasonable to expect that strong magnetic Fe-Cr interaction might influence the Curie temperature of the alloy. To investigate this, finite-temperature Monte Carlo simulations were performed for all the compounds, with 1024 Cr atoms (6.25 at. %) added to the simulation cell, again randomly replacing Fe, Ni, or both Fe and Ni (512 atoms of each species). Figure 7 shows the temperature dependence of the total magnetic moment. For all the three compounds, the addition of Cr results in reduction of the total magnetic moment. Cr also changes the Curie temperature of all the alloys, but it is in $Fe_3Ni$ where this change is most dramatic. In the $Fe_3Ni$ compound the replacement of Ni by randomly placed Cr atoms increases the temperature of the magnetic transition from ~500 K to well over 700 K. The replacement of both Fe and Ni atoms with Cr also increases the $T_C$, whereas the replacement of Fe alone by Cr does not have such an effect (Figure 7a). In the $L1_0$ FeNi compound, the replacement of Ni by Cr results in the Curie temperature increase of less than 100K, whereas in $FeNi_3$ all the possible substitutions of atoms with Cr result in the decrease of the $T_C$. We interpret this finding as related to the ferromagnetic first, third and fourth nearest neighbor Ni-Cr interactions (see Table 1), which are weaker than ferromagnetic Ni-Ni interactions. As a result, in Ni-rich alloys the replacement of Ni with Cr results in the decrease of the $T_C$, as opposed to the case of Fe-rich alloys. We note that the effect of Curie temperature increase has already been found experimentally in bcc Fe-Cr alloys with small (below 10 at. %) chromium content [22,27], and it was explained by strong antiferromagnetic Fe-Cr interactions [22]. An experimental study of disordered fcc $(FeNi)_{1-x}Cr_x$ alloys for $x$ = 0, 5, 10, and 15 at. % was performed in [28] and showed a substantial decrease of the Curie temperature as a function of increasing chromium content (from almost 800 K for $x$ = 0 down to under 200 K for $x$ = 15 at. %). The authors of Ref. [28] also noted some appreciable variation of magnetic ordering temperatures for 10 and 15 at. % Cr alloys. Comparison of their results for $T_C$ with our simulations performed for the ordered $L1_0$ FeNi compound and completely random Fe-Ni alloys with Cr replacing both iron and nickel suggests a transition from ordered FeNi to random Fe-Ni-Cr alloy with increasing chromium content, which probably affects experimental observations. In view of this it may be worth investigating magnetic properties of ordered and random FeNi systems with controlled Cr replacement of one or the other alloy components, or both of them.

We also explored magnetic properties of ternary ordered compound $Fe_2NiCr$. The crystal structure of $Fe_2NiCr$ is similar to $L1_0$ FeNi, where one of the Ni atoms in a unit cell is replaced by Cr. This structure was extensively studied using *ab initio* methods in Ref. [4] and was found to have a lower value of the enthalpy of mixing than all the experimentally known intermetallic phases of fcc Fe-Ni-Cr alloys. An initial MCE investigation of this alloy was performed in [4]. Because of its significance, we simulated its properties again using the improved MCE model developed here. Results of simulations of magnetic ground states of random and ordered $Fe_2NiCr$ are summarized in Table 4, together with results of DFT calculations [4]. Ordered $Fe_2NiCr$ was found to have an almost exactly collinear magnetic structure. While our simulations show that a random mixture with atomic content $Fe_{50}Ni_{25}Cr_{25}$ is almost completely antiferromagnetic (the average magnetic moment is predicted to be 0.025 $\mu_B$ per atom with no collinearity constraint applied, and 0.206 $\mu_B$ per atom if simulations are constrained to be collinear), an ordered structure with the same composition has large nonzero total magnetic moment, with Cr moments being anti-ferromagnetically ordered with respect to the Fe moments. The moments of Ni atoms in all the calculations, including *ab initio* studies, are significantly smaller than atomic moments in pure fcc Ni (0.575 $\mu_B$ predicted by MCE simulations as compared to 0.605 $\mu_B$ found in experimental observations [22]). Ni moments are aligned ferromagnetically with respect to the Fe moments. We note that the reduction of Ni moments in ternary Fe-Ni-Cr alloys, compared to pure Ni and binary Fe-Ni alloys, is a generic phenomenon found both in *ab initio* calculations (see Tables 3 and 4) and in MCE simulations. Extensive DFT calculations summarized in [4] show that magnetic moments on Ni sites in alloys containing more than 33 at. % Cr are close to zero, and that Cr-Ni alloys containing over 20 at. % Cr are non-magnetic. Magnetic moments of the ordered $Fe_2CrNi$ alloy and each of its components are plotted in Figure 8 as functions of temperature. The alloy remains magnetic until fairly high temperatures. Simulations performed using the current MCE Hamiltonian predict the Curie temperature close to 1050-1100 K, which is slightly higher than the value of ~1000 K found using an earlier parameterization [4]. The effect of transition temperature increase compared to pure Ni and binary FeNi and $FeNi_3$ alloys is similar to the one observed in fcc Fe-Ni, where the chemically ordered $FeNi_3$ compound has higher Curie temperature than pure Ni. In relation to the ternary $Fe_2CrNi$ alloy, we again attribute the high stability of its magnetically ordered configuration to strong antiferromagnetic coupling between Fe and Cr atoms.

## V. Conclusions

This paper describes a new Magnetic Cluster Expansion model and its application to a technologically relevant ternary magnetic Fe-Ni-Cr fcc alloy. Despite the fact that the MCE formalism involves several approximations, for example the model neglects the environmental dependence of the Landau on-site terms, the low temperature predictions derived from the model agree well with DFT data. We are also able to explore high temperature magnetic properties of the alloys, by performing Monte Carlo simulations for both random and ordered alloy configurations. Strong antiferromagnetic Fe-Cr interaction is responsible for that during alloying with Cr, chromium atoms prefer replacing Ni atoms in all the ordered Fe-Ni compounds. The replacement of Ni atoms by Cr also increases the Curie temperature of Fe-rich ordered alloys. The interplay between chemical and magnetic degrees of freedom is responsible for the very high Curie temperature of ordered $Fe_2CrNi$ alloys, somewhat similar to the case of bcc Fe-Cr alloys [20] where the Curie temperature is maximum at 6 at. % Cr. MCE predictions agree very well with the available experimental data and *ab initio* calculations performed in the collinear magnetic approximation [3]. This shows that MCE Hamiltonian-based Monte Carlo simulations can be successfully applied to ternary magnetic alloys exhibiting ferromagnetic and antiferromagnetic properties. Further improvement in the accuracy of MCE models for multi-component magnetic alloys can likely be achieved through the use of larger *ab initio* DFT databases generated using a constrained non-collinear magnetic methodology [29,30].

**Acknowledgements**

This work was part-funded by the EuroFusion Consortium, and has received funding from Euratom research and training programme 2014-2018 under grant agreement number No 633053, and funding from the RCUK Energy Programme (Grant Number EP/I501045). The views and opinions expressed herein do not necessarily reflect those of the European Commission. To obtain further information on the data and models underlying this paper please contact PublicationsManager@ccfe.ac.uk. This work was also part-funded by the United Kingdom Engineering and Physical Sciences Research Council via a programme grant EP/G050031. This work was also partly funded by the Accelerated Metallurgy Project, which is co-funded by the European Commission in the 7th Framework Programme (Contract NMP4-LA-2011-263206), by the European Space Agency and by the individual partner organizations. DNM would like to acknowledge the Juelich supercomputer center for the provision of High-Performances Computer for Fusion (HPC-FF) facilities as well as the International Fusion Energy Research Centre (IFERC) for

Table 1. Non-magnetic interaction parameters $I_{ij}$ and magnetic Heisenberg interaction parameters $Y_{ij}$ (meV) derived using a fitting procedure described in [3].

|       | 1st NN $I_{ij}$ | 1st NN $Y_{ij}$ | 2nd NN $I_{ij}$ | 2nd NN $Y_{ij}$ | 3rd NN $I_{ij}$ | 3rd NN $Y_{ij}$ | 4th NN $I_{ij}$ | 4th NN $Y_{ij}$ |
|-------|-----------|-----------|-----------|-----------|-----------|-----------|-----------|-----------|
| Fe-Fe | 1.856364  | -0.793072 | 10.741989 | -10.827175| -0.405778 | 0.546719  | -2.047610 | 2.305911  |
| Fe-Ni | -2.710858 | -3.805516 | -12.447877| -1.487362 | 0.131012  | -0.530692 | -3.104198 | 0.136000  |
| Fe-Cr | -6.627640 | 5.778819  | -4.571750 | 0.488416  | -4.791461 | -0.309644 | 6.560627  | -0.366602 |
| Ni-Ni | 1.132506  | -13.153009| 0.006062  | 7.227536  | 11.972890 | -5.604799 | 2.557930  | -6.744045 |
| Ni-Cr | -0.419121 | -5.501130 | 7.440212  | 0.692985  | -11.923519| -0.820111 | -7.328073 | -0.629514 |
| Cr-Cr | -3.933508 | -0.741406 | -4.550795 | -0.011726 | -4.898933 | 0.416839  | 2.322535  | 0.296039  |

Table 2. The on-site Landau expansion terms (in meV units) entering the MCE Hamiltonian (1).

|   | Fe | Ni | Cr |
|---|---|---|---|
| $A$ | -0.99016 | 30.37460 | -3.47938 |
| $B$ | 29.05331 | 455.69 | 5.226 |
| $C$ | -6.49401 | -138.05 | |
| $D$ | 0.42817 | 18.8 | |

Table 3. Magnetic moments per atom computed for several SQS structures using the Magnetic Cluster Expansion model (with and without the collinearity constraint applied) and DFT ($\mu_B$).

|   |   | MCE (non-collinear) | MCE (collinear) | DFT |
|---|---|---|---|---|
| 1 | $M_{total}$ | 0.674 | 1.221 | 1.147 |
|   | $M_{Fe}$ | 1.244 | 2.202 | 2.016 |
|   | $M_{Ni}$ | 0.294 | 0.415 | 0.336 |
|   | $M_{Cr}$ | 1.372 | 2.155 | 1.776 |
| 2 | $M_{total}$ | 0.727 | 1.154 | 1.123 |
|   | $M_{Fe}$ | 1.211 | 2.081 | 1.957 |
|   | $M_{Ni}$ | 0.331 | 0.433 | 0.348 |
|   | $M_{Cr}$ | 0.942 | 2.060 | 1.682 |
| 3 | $M_{total}$ | 0.493 | 0.386 | 1.159 |
|   | $M_{Fe}$ | 0.919 | 0.932 | 2.036 |
|   | $M_{Ni}$ | 0.224 | 0.041 | 0.321 |
|   | $M_{Cr}$ | 1.040 | 1.433 | 1.585 |
| 4 | $M_{total}$ | 0.836 | 1.124 | 1.045 |
|   | $M_{Fe}$ | 1.600 | 2.111 | 1.844 |
|   | $M_{Ni}$ | 0.422 | 0.454 | 0.399 |
|   | $M_{Cr}$ | 1.748 | 2.142 | 1.530 |
| 5 | $M_{total}$ | 0.641 | 1.200 | 1.115 |
|   | $M_{Fe}$ | 1.185 | 2.222 | 1.952 |
|   | $M_{Ni}$ | 0.332 | 0.436 | 0.397 |
|   | $M_{Cr}$ | 1.185 | 2.136 | 1.557 |

Table 4. Magnetic moments per atom computed for the ground states of random (without and with the collinearity constraint) and ordered $Fe_2NiCr$ alloy using the Magnetic Cluster Expansion model, and for the ordered $Fe_2NiCr$ compound using DFT [4] ($\mu_B$).

|  | Random | | Ordered | |
| --- | --- | --- | --- | --- |
|  | Non-collinear MCE | Collinear MCE | MCE (collinear) | DFT [4] |
| $M_{total}$ | 0.025 | 0.206 | 0.921 | 0.471 |
| $M_{Fe}$ | 0.072 | 0.552 | 2.715 | 2.085 |
| $M_{Ni}$ | 0.013 | 0.130 | 0.381 | 0.152 |
| $M_{Cr}$ | 0.056 | 0.410 | 2.131 | 2.437 |

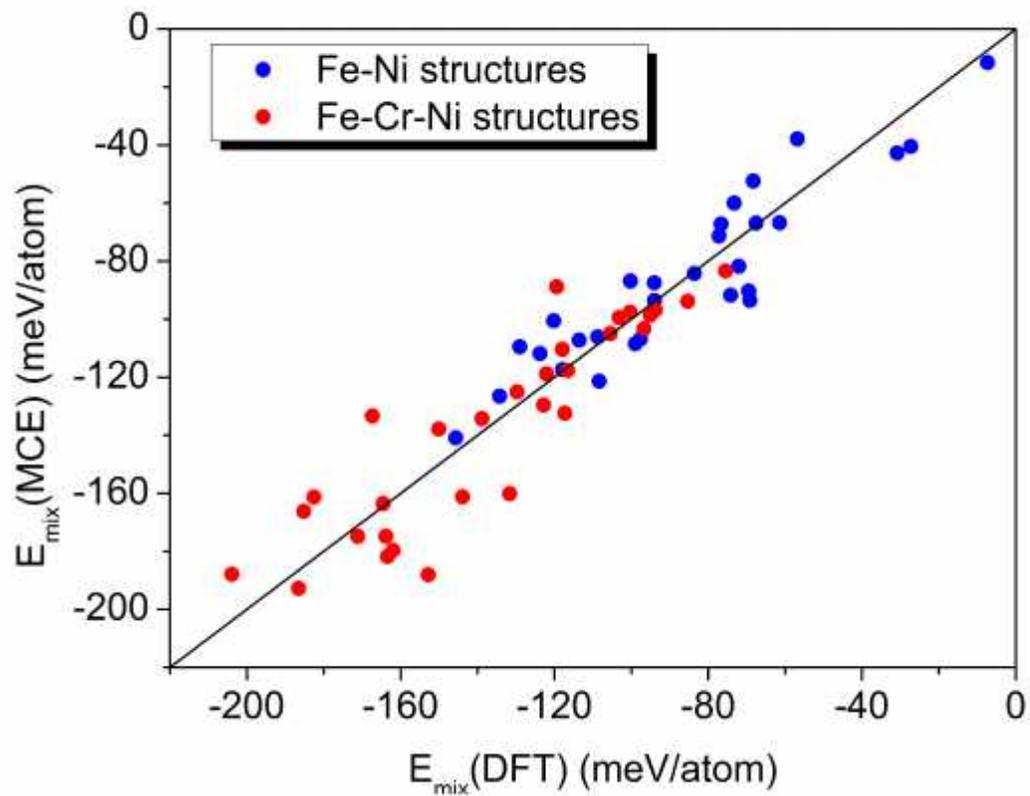

Figure 1. Comparison between DFT and MCE energies of mixing for alloy configurations used for fitting the MCE Hamiltonian. Data points for Fe-Ni and Fe-Cr-Ni alloys are shown using different colours.

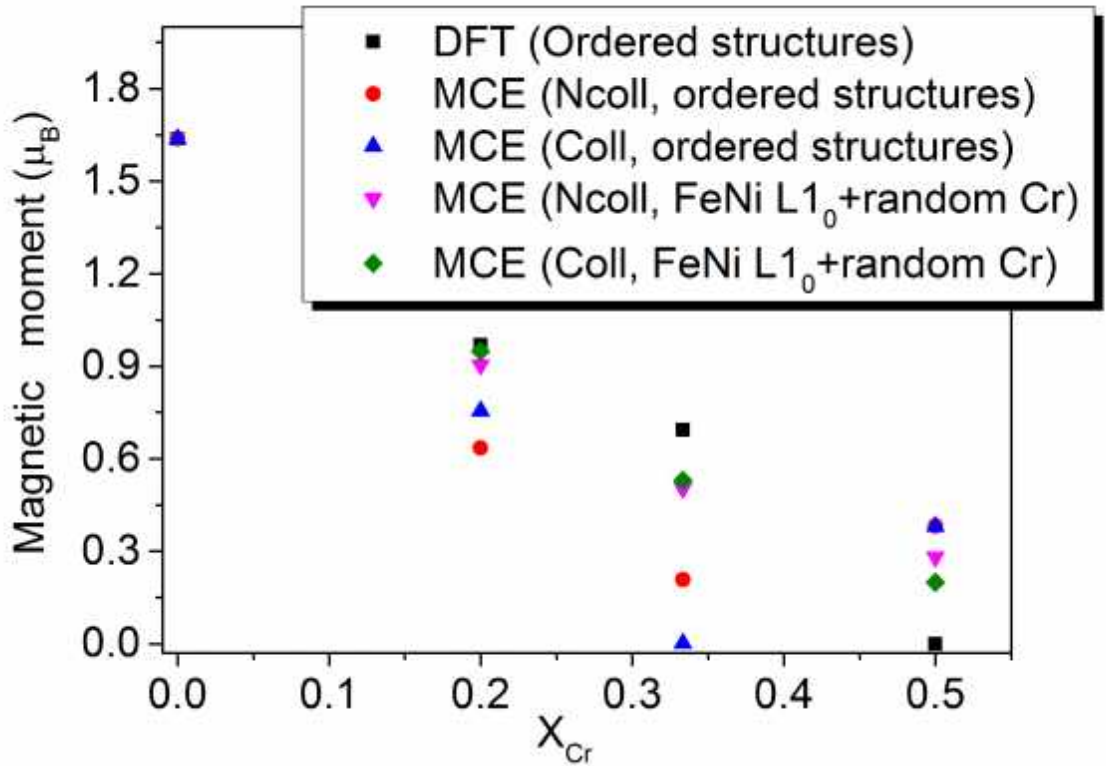

Figure 2. Total magnetic moment per atom in $(Fe_{0.5}Ni_{0.5})_{1-x}Cr_x$ alloys computed assuming ordered and disordered alloy configurations, with (Coll) and without (Ncoll) the collinearity constraint applied. DFT results (black squares) are shown for comparison.

Figure 3. Magnetic moment of random Fe-Ni-Cr mixture ($\mu_B$) without (a) and with (b) collinearity constraint applied.

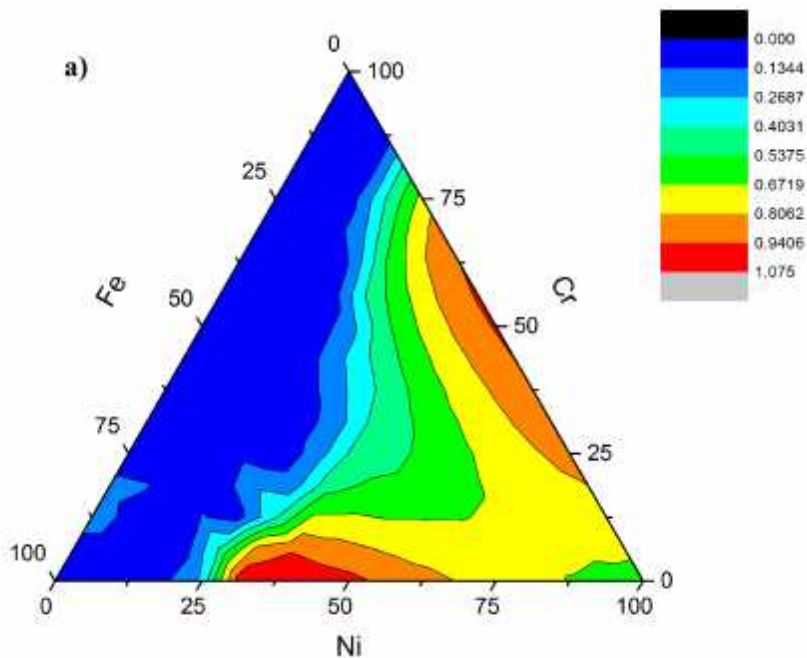

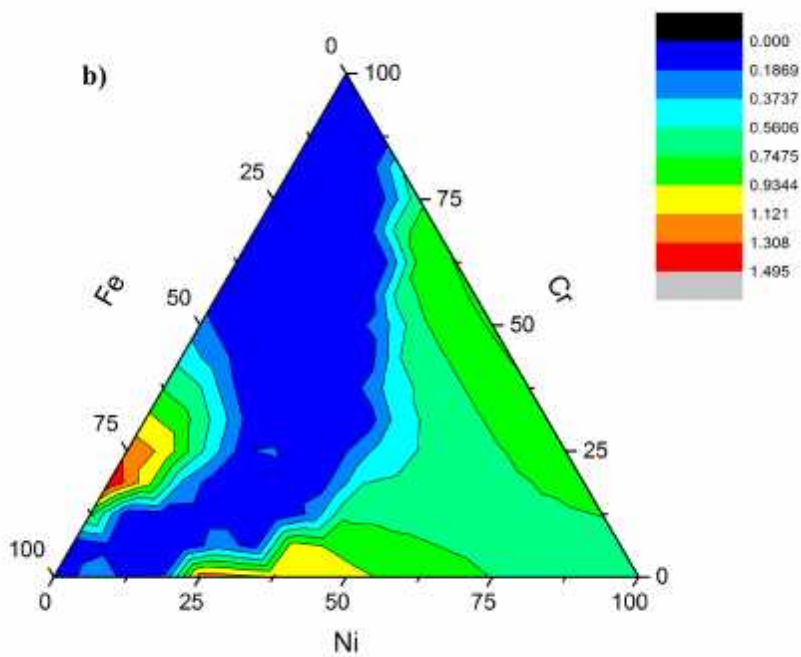

Figure 4. Enthalpy of mixing (a) and the formation enthalpy (b) of random Fe-Ni-Cr mixtures (meV/atom) simulated with no collinearity constraint applied.

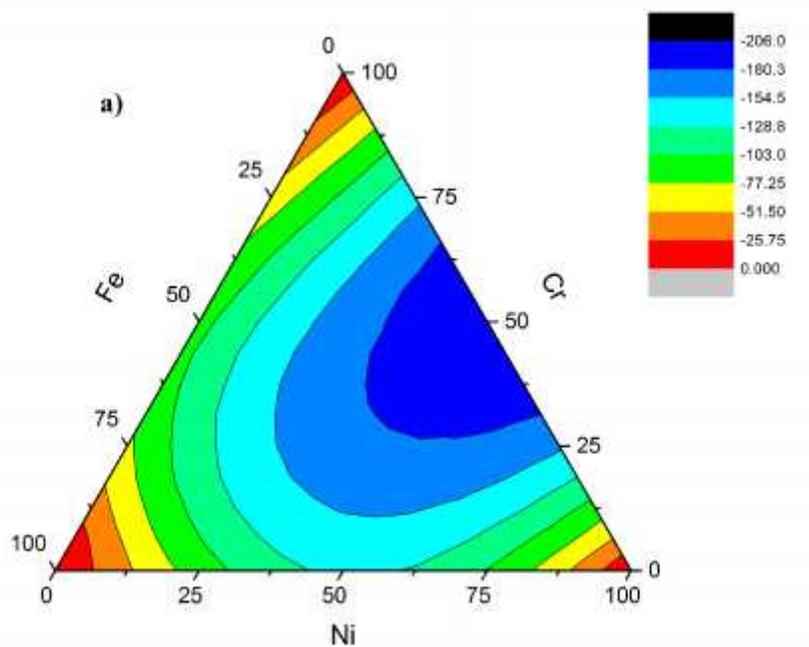

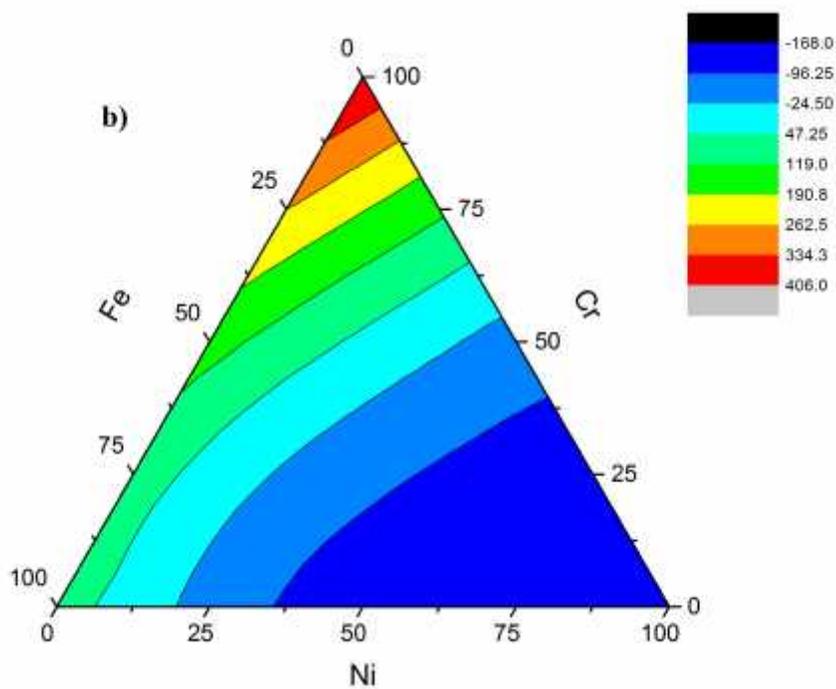

Figure 5. Magnetic moment of random Fe-Ni-Cr mixtures ($\mu_B$) at T=500 K.

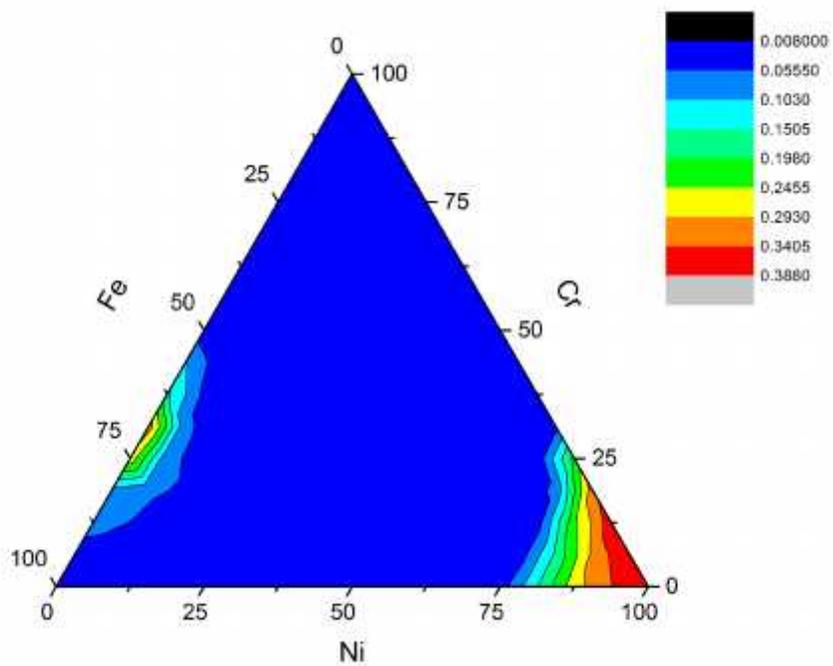

Figure 6. Energy of ordered Fe-Ni fcc structures with randomly distributed Cr replacing Fe, Ni, or both Fe and Ni atoms. The curves refer to the following ordered alloy structures: $Fe_3Ni$ $L1_2$ (a), FeNi $L1_0$ (b), $FeNi_3$ $L1_2$ (c).

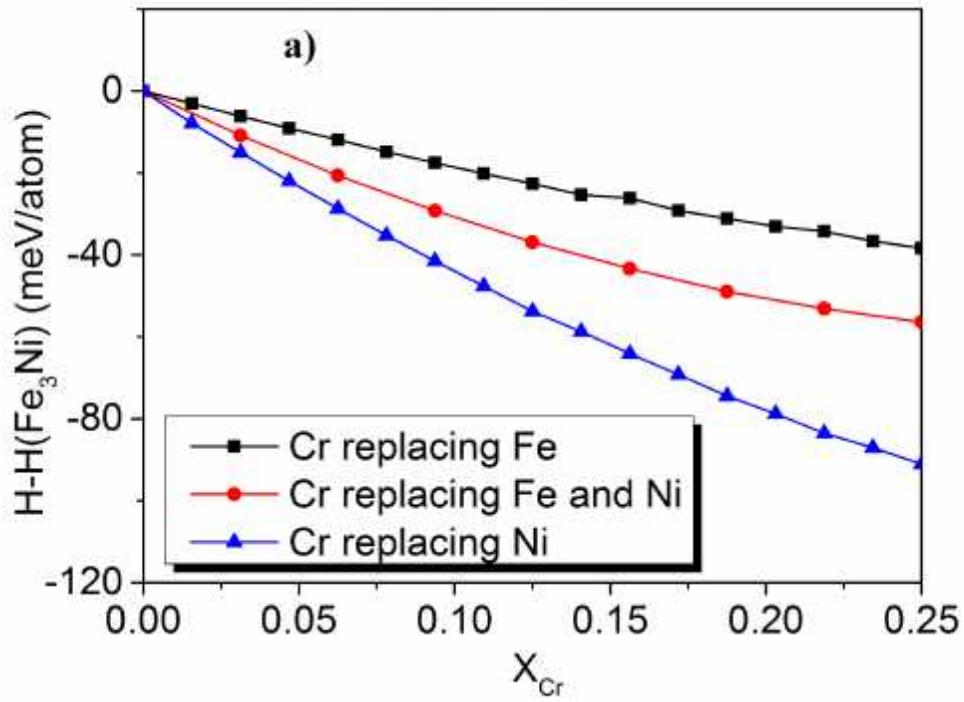

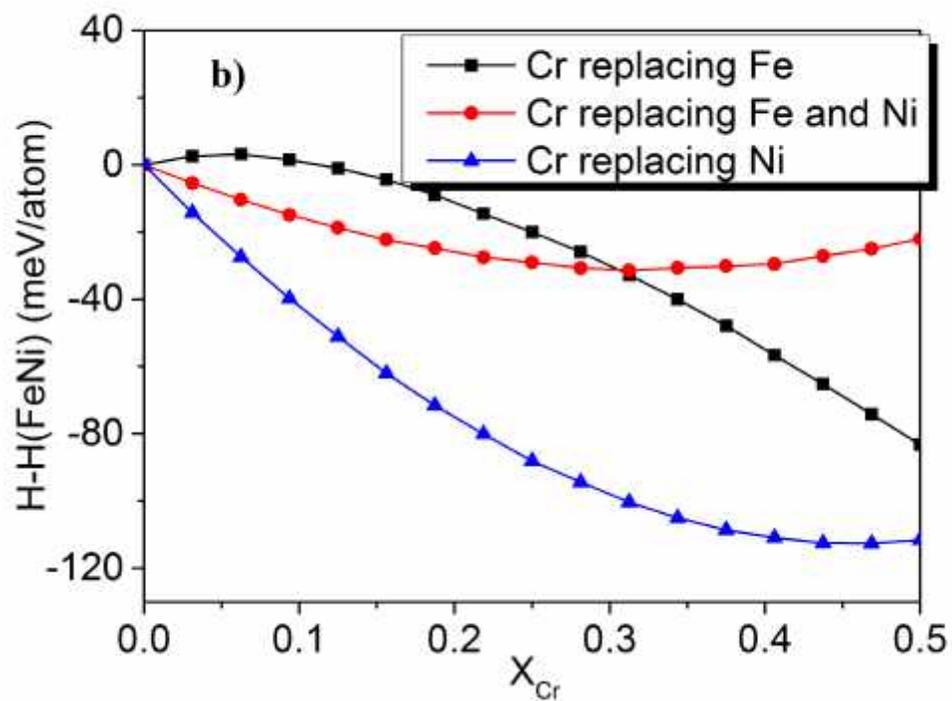

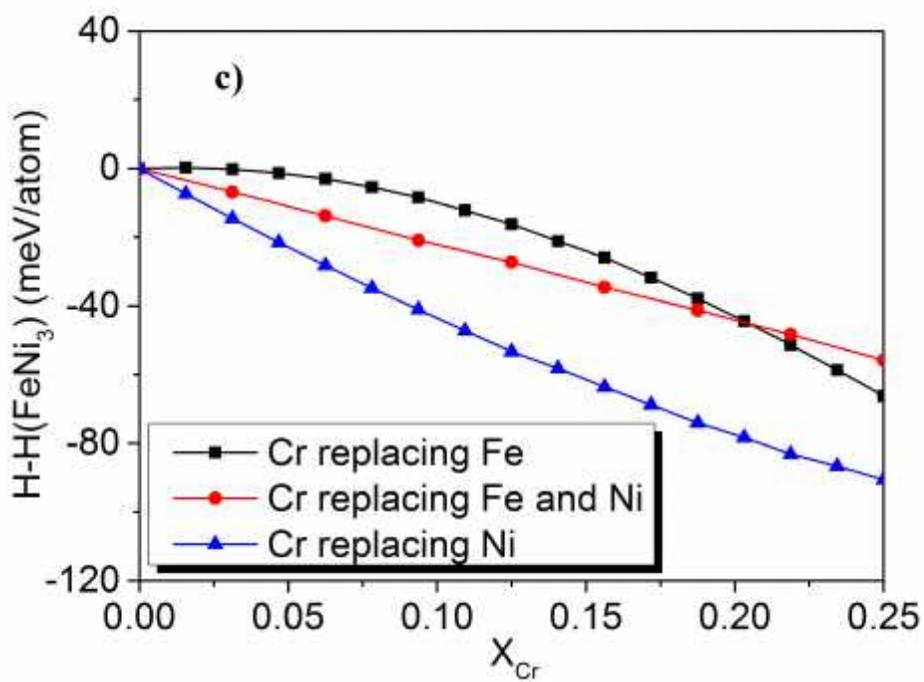

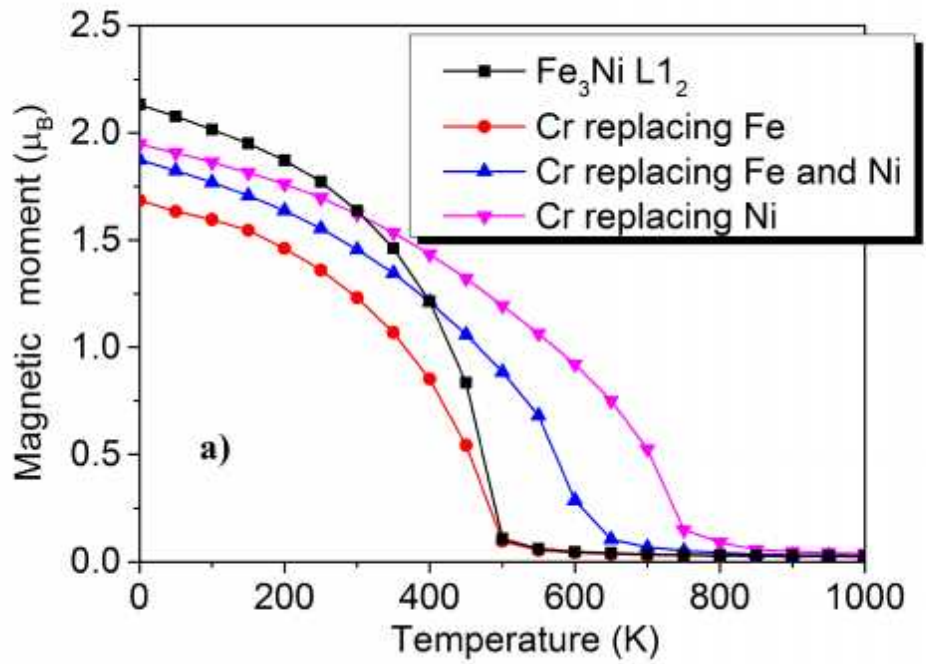

Figure 7. Magnetic moments of ordered Fe-Ni fcc alloys containing randomly distributed 6.25 at. % Cr atoms replacing Fe, Ni, or both Fe and Ni atoms. The curves refer to the following alloy structures: $Fe_3Ni$ $L1_2$ (a), FeNi $L1_0$ (b), $FeNi_3$ $L1_2$ (c). The moments computed for ordered Fe-Ni alloys with no Cr present are shown for comparison.

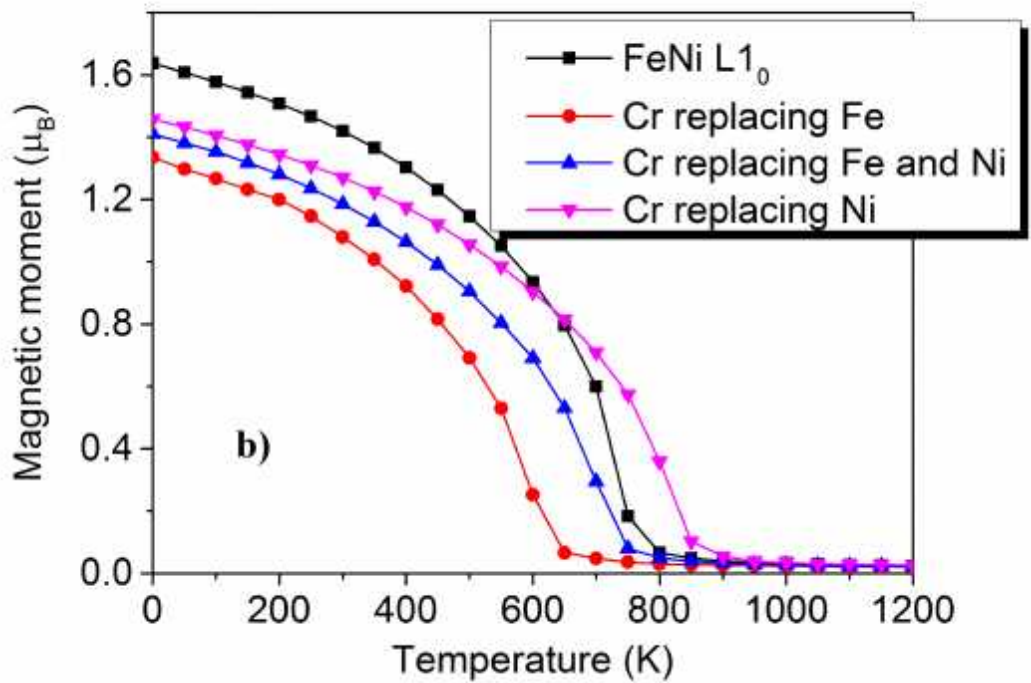

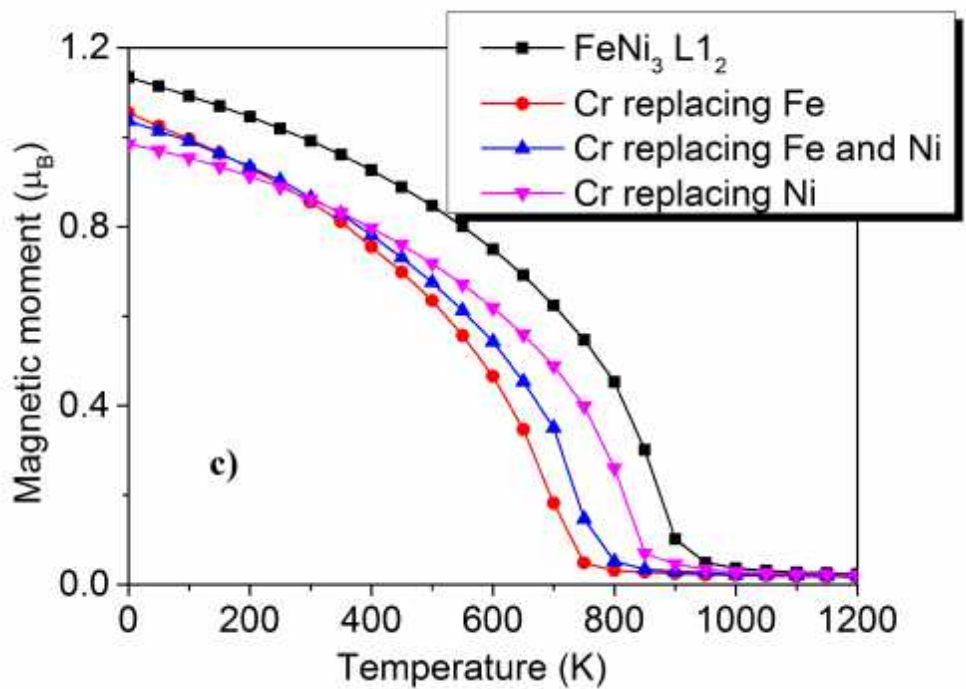

Figure 8. Temperature dependence of the total magnetic moment of ordered Fe$_2$CrNi alloy, and the moments of atoms forming the alloy.

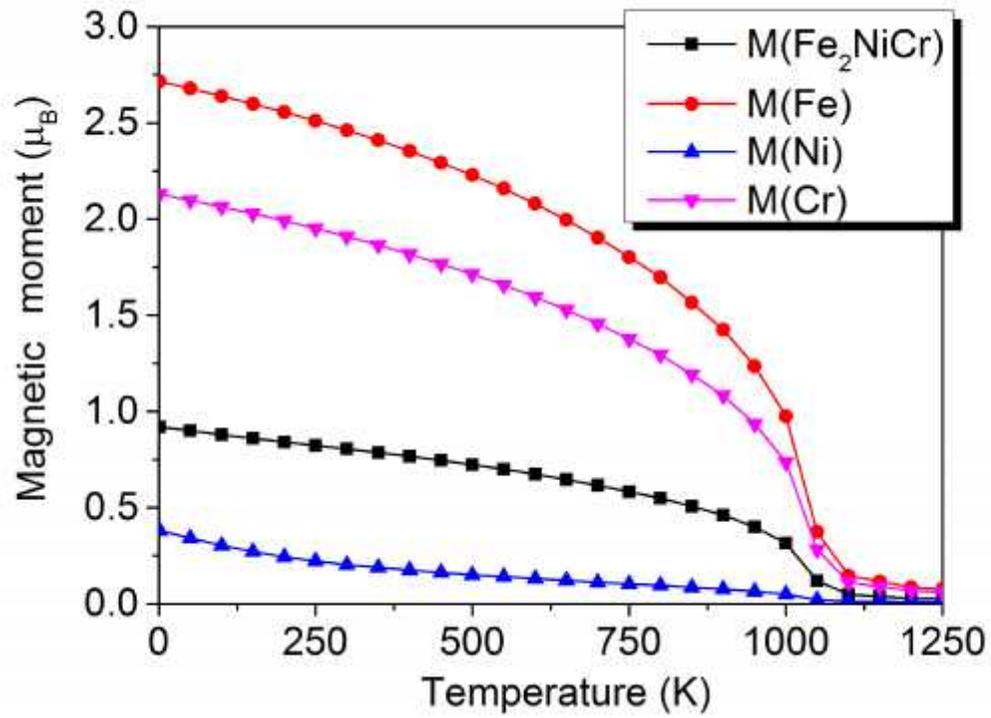